\documentclass[preprint,12pt]{aastex}

\shortauthors{Winn et al.~2006}
\shorttitle{Spin-Orbit Alignment in HD~189733}

\begin{document}

% ------------------------------------------------------------------------
% New commands
%
\def\ltsima{$\; \buildrel < \over \sim \;$}
\def\lsim{\lower.5ex\hbox{\ltsima}}
\def\gtsima{$\; \buildrel > \over \sim \;$}
\def\gsim{\lower.5ex\hbox{\gtsima}}
\def\lam{\lambda=-1\fdg4 \pm 1\fdg1}
                                                                                          
% -------------------------------------------------------------------------
%

\bibliographystyle{apj}

\title{
Measurement of the Spin-Orbit Alignment in\\
the Exoplanetary System HD~189733$^1$
}

\author{
Joshua N.\ Winn\altaffilmark{2},
John Asher Johnson\altaffilmark{3},
Geoffrey W.\ Marcy\altaffilmark{3},
R. Paul Butler\altaffilmark{4},\\
Steven S. Vogt\altaffilmark{5},
Gregory W.\ Henry\altaffilmark{6},
Anna Roussanova\altaffilmark{2},
Matthew J. Holman\altaffilmark{7},\\
Keigo Enya\altaffilmark{8},
Norio Narita\altaffilmark{9},
Yasushi Suto\altaffilmark{9},
Edwin L.\ Turner\altaffilmark{10}
}

\altaffiltext{1}{Data presented herein were obtained at the W.M.~Keck
  Observatory, which is operated as a scientific partnership among the
  California Institute of Technology, the University of California,
  and the National Aeronautics and Space Administration, and was made
  possible by the generous financial support of the W.~M.~Keck
  Foundation.}

\altaffiltext{2}{Department of Physics, and Kavli Institute for
  Astrophysics and Space Research, Massachusetts Institute of
  Technology, Cambridge, MA 02139, USA}

\altaffiltext{3}{Department of Astronomy, University of California,
  Mail Code 3411, Berkeley, CA 94720, USA}

\altaffiltext{4}{Department of Terrestrial Magnetism, Carnegie
  Institution of Washington, 5241 Broad Branch Road NW, Washington
  D.C. USA 20015-1305}

\altaffiltext{5}{UCO/Lick Observatory, University of California at
  Santa Cruz, Santa Cruz CA USA 95064}

\altaffiltext{6}{Center of Excellence in Information Systems,
  Tennessee State University, 3500 John A.\ Merritt Blvd., Box 9501,
  Nashville, TN 37209, USA}

\altaffiltext{7}{Harvard-Smithsonian Center for Astrophysics, 60
  Garden Street, Cambridge, MA 02138, USA}

\altaffiltext{8}{Institute of Space and Astronautical Science, Japan
  Aerospace Exploration Agency, 3-1-1, Yoshinodai, Sagamihara,
  Kanagawa, 229-8510, Japan}

\altaffiltext{9}{Department of Physics, The University of Tokyo, Tokyo
  113-0033, Japan}

\altaffiltext{10}{Princeton University Observatory, Peyton Hall,
  Princeton, NJ 08544, USA}

\begin{abstract}

  We present spectroscopy of a transit of the exoplanet HD~189733b. By
  modeling the Rossiter-McLaughlin effect (the anomalous Doppler shift
  due to the partial eclipse of the rotating stellar surface), we find
  the angle between the sky projections of the stellar spin axis and
  orbit normal to be $\lam$. This is the third case of a ``hot
  Jupiter'' for which $\lambda$ has been measured. In all three cases
  $\lambda$ is small, ruling out random orientations with 99.96\%
  confidence, and suggesting that the inward migration of hot Jupiters
  generally preserves spin-orbit alignment.

\end{abstract}

\keywords{planetary systems --- planetary systems: formation ---
  stars:~individual (HD~189733) --- stars:~rotation}

\section{Introduction}

A primary reason to study planets of other stars is to learn how
typical (or unusual) are the properties of the Solar system. For
example, the nearly circular orbits of Solar system planets were once
considered normal, but we now know that eccentric orbits of Jovian
planets are common (see, e.g., Halbwachs, Mayor, \& Udry 2005; or
Fig.~3 of Marcy et al.~2005). Likewise, gas giants were once thought
to inhabit only the far reaches of planetary systems, an assumption
that was exploded by the discovery of ``hot Jupiters'' (Mayor \&
Queloz 1995, Butler et al.~1997). This inspired theoretical work on
planetary migration mechanisms that can deliver Jovian planets to such
tight orbits (as recently reviewed by Thommes \& Lissauer 2005 and
Papaloizou \& Terquem 2006).

Another striking pattern in the Solar system is the close alignment
between the planetary orbits and the Solar spin axis. The orbit
normals of the 8 planets are within a few degrees of one another (Cox
et al.~2000, p.~295), and the Earth's orbit normal is only 7~degrees
from the Solar spin axis (Beck \& Giles 2005, and references
therein). Presumably this alignment dates back 5~Gyr, when the Sun and
planets condensed from a single spinning disk. Whether or not this
degree of alignment is universal is unknown. For hot Jupiters in
particular, one might wonder whether migration enforces or perturbs
spin-orbit alignment.

For exoplanets, the angle between the stellar spin axis and planetary
orbit normal (as projected on the sky) can be measured via the
Rossiter-McLaughlin (RM) effect: the spectral distortion observed
during a transit due to stellar rotation. The planet hides some of the
velocity components that usually contribute to line broadening,
resulting in an ``anomalous Doppler shift'' (for the theory, see Ohta
et al.~2005, Gim\'enez 2006, or Gaudi \& Winn~2006).

Observations of the exoplanetary RM effect have been reported for
HD~209458 (Bundy \& Marcy 2000, Queloz et al.~2000, Winn et al.~2005,
Wittenmyer et al.~2005) and HD~149026 (Wolf et al.~2006). Here we
report observations of the RM effect for HD~189733. This system,
discovered by Bouchy et al.~(2005), consists of a K dwarf with a
transiting Jovian planet ($M_P = 1.15$~$M_{\rm Jup}$) in a 2.2~day
orbit. Our observations are presented in \S~2, our model in \S~3, and
our results in \S~4, followed by a brief summary and discussion.

\section{Observations}

We observed the transit of UT~2006~August~21 with the Keck~I 10m
telescope and the High Resolution Echelle Spectrometer (HIRES; Vogt et
al.~1994) following the usual protocols of the California-Carnegie
planet search, as summarized here. We employed the red cross-disperser
and placed the I$_2$ absorption cell into the light path to calibrate
the instrumental response and the wavelength scale. The slit width was
$0\farcs85$ and the typical exposure time was 3~minutes, giving a
resolution of 70,000 and a signal-to-noise ratio (SNR) of
300~pixel$^{-1}$. We obtained 70 spectra over 7.5~hr bracketing the
predicted transit midpoint. To these were added 16 spectra that had
been obtained by the California-Carnegie group at random orbital
phases.

We determined the relative Doppler shifts with the algorithm of Butler
et al.~(1996). We estimated the measurement uncertainties based on the
scatter in the solutions for each 2~\AA~section of the spectrum. For
the spectra obtained on 2006~Aug~21 the typical measurement error was
0.8~m~s$^{-1}$, while for the other 16 spectra the error was
$\approx$1.3~m~s$^{-1}$. The data are given in Table~1 and plotted in
Fig.~1, with enlarged error bars to account for the intrinsic velocity
noise of the star (see \S~3).

We also needed accurate photometry to pin down the planetary and
stellar radii and the orbital inclination. We observed the transit of
UT~2006~Jul~21 with KeplerCam on the 1.2m telescope at the Fred L.\
Whipple Observatory on Mt.\ Hopkins, Arizona. We used the SDSS
$z$~band filter and an exposure time of 30 seconds. After bias
subtraction and flat-field division, we performed aperture photometry
of HD~189733 and 14 comparison stars. The light curve of each
comparison star was normalized to have unit median, and the mean of
these normalized light curves was taken to be the comparison
signal. The light curve of HD~189733 was divided by the comparison
signal, and corrected for residual systematic effects by dividing out
a linear function based on the out-of-transit data. The light curve is
plotted in the top panel of Fig.~1.

\begin{figure}[p]
\epsscale{0.65}
\plotone{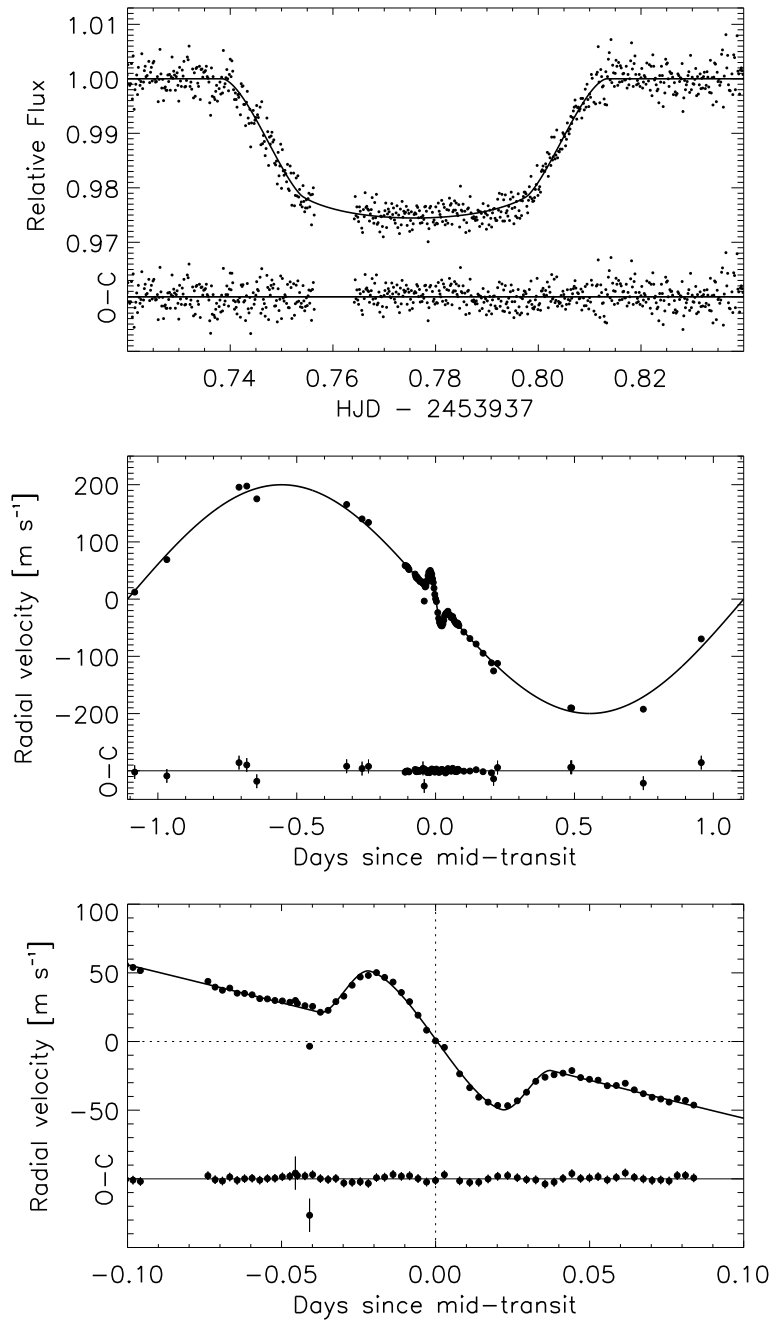}
\caption{
Photometry and spectroscopy of HD~189733.
The top panel shows $z$ band photometry during a transit, along with the
best-fitting model (solid line).
The middle panel show the radial velocities
as a function of orbital phase (expressed in days), along with the model
(solid line).
The bottom panel is a close-up near the mid-transit time.
In all cases, the residuals
(observed$-$calculated) are plotted beneath the data.
\label{fig:1}}
\end{figure}

\section{The Model}

We fitted the fluxes and radial velocities with a parameterized model
based on a star and planet in a circular orbit about the center of
mass.\footnote{A circular orbit is a reasonable simplifying assumption
  because of the effects of tidal circularization (see, e.g., Rasio et
  al.~1996, Trilling et al.~2000, Dobbs-Dixon et al.~2004).} To
calculate the relative flux as a function of the projected separation
of the planet and the star, we assumed the limb darkening law to be
quadratic and employed the analytic formulas of Mandel \& Agol~(2002)
to compute the integral of the intensity over the unobscured portion
of the stellar disk. We fixed the limb-darkening coefficients at the
values $u_1=0.320$, $u_2=0.267$, based on the calculations of
Claret~(2004).

To calculate the anomalous Doppler shift, we used the technique of
Winn et al.~(2005): we simulated RM spectra with the same data format
and noise characteristics as the actual data, and determined the
Doppler shifts using the same algorithm used on the actual data. The
simulations were based on a ``template'' spectrum representing the
emergent spectrum from a small portion of the photosphere. We scaled
the template spectrum in flux by $\epsilon$ and shifted it in velocity
by $v_p$, representing the spectrum of the occulted portion of the
stellar disk. We subtracted this spectrum from a
rotationally-broadened version of the template spectrum (broadened to
3~km~s$^{-1}$ to mimic the disk-integrated spectrum of HD~189733), and
then ``measured'' the anomalous Doppler shift $\Delta v$. This was
repeated for a grid of $\{\epsilon, v_p\}$, and a polynomial function
was fitted to the resulting surface.

The template spectrum should be similar to that of HD~189733 but
without significant rotational broadening. We tried three different
choices: two empirical spectra, and one theoretical spectrum. The two
empirical spectra were Keck/HIRES spectra (SNR~$\approx~800$,
$R\approx~10^5$) of HD~3561 (G3~{\sc v}, $v\sin i_S=1.2\pm
0.5$~km~s$^{-1}$) and HD~3765 (K2~{\sc v}, $0.0\pm
0.5$~km~s$^{-1}$). The former is 200~K hotter than HD~189733, while
the latter is more metal-rich. The theoretical spectrum, with a
resolution of 250,000, was taken from Coelho et al.~(2005) for a
non-rotating star with $T_{\rm eff}=5000$~K, $\log g=4.5$, and
[Fe/H]~$=0.0$.

For each choice of the template spectrum, we derived the relation
between $\Delta v$, $\epsilon$, and $v_p$, and optimized the model as
described below. With one exception, the results did not depend
significantly on the choice of template spectrum (in the sense that
measurement errors caused much larger uncertainties). The single
exception was $v\sin i_S$, for which the results differed as much as
3\%. For our final analysis, we used the relation
\begin{equation}
\Delta v = -\epsilon~v_p\left[1.252 - 0.351 \left( \frac{v_p}{{\rm 3~km~s}^{-1}} \right)^2 \right]
\end{equation}
derived from the empirical templates, but we also included an extra
error term of 6\% in $v\sin i_S$ as a conservative estimate of the
systematic error. In summary, the projected separation of the planet
and the star determines the transit depth $\epsilon$ and the
sub-planet rotation velocity\footnote{The sub-planet velocity is the
  projected rotation velocity of the portion of the star hidden by the
  planet, and is calculated assuming no differential rotation, an
  assumption justified by Gaudi \& Winn (2006).} $v_p$, and then
Eq.~(1) is used to calculate the anomalous Doppler shift.

The fitting statistic was
\begin{equation}
\chi^2 =
\sum_{j=1}^{86}
\left[
\frac{v_j({\mathrm{obs}}) - v_j({\mathrm{calc}})}{\sigma_{v,j}}
\right]^2
+ 
\sum_{j=1}^{752}
\left[
\frac{f_j({\mathrm{obs}}) - f_j({\mathrm{calc}})}{\sigma_{f,j}}
\right]^2
+
\left( \frac{\Delta\gamma}{{\rm 12~m~s^{-1}}} \right)^2
+
\left( \frac{M_S/M_\odot - 0.82}{0.03} \right)^2
,
\end{equation}
where $f_j$(obs) is the flux observed at time $j$, $\sigma_{f,j}$ is
the corresponding uncertainty, and $f_j$(calc) is the calculated
value. A similar notation applies to the velocities. The last two
terms are {\it a priori} constraints explained below.

It is important for $\sigma_{f,j}$ and $\sigma_{v,j}$ to include not
only measurement errors but also any unmodeled systematic errors. To
account for systematic errors in the photometry, we increased the
Poisson estimates of the errors by a factor of 1.2, at which point
$\chi^2/N_{\rm DOF} = 1$ when fitting only the fluxes. Determining the
appropriate weights for the velocities was more complex. HD~189733 is
chromospherically active and should exhibit velocity noise
(``photospheric jitter'') with an amplitude of 11~m~s$^{-1}$ according
to the empirical relations of Wright~(2005). However, the time scale
of the noise cannot be predicted as easily. Noise from spots or plages
would occur on the time scale of the rotation period
($\approx$13~days), while noise from oscillations and flows occurs on
shorter time scales.

We took the following approach. First, we fitted only the 16
velocities obtained sporadically prior to 2006~Aug~21, and found the
root-mean-squared (RMS) residual to be 12~m~s$^{-1}$, in agreement
with the Wright~(2005) relations. Therefore, for fitting purposes, we
inflated the error bars $\sigma_{v,j}$ of those 16 velocities to
12~m~s$^{-1}$. Second, we fitted only the 44 {\it out-of-transit}
velocities measured on 2006~Aug~21, and found the RMS residual to be
1.5~m~s$^{-1}$. In addition, there were correlations in the residuals
on a time scale of $\sim$15~minutes ($\sim$4 data points). The
correlations effectively reduce the number of independent data points
by 4, or equivalently, they double the error per point. Therefore, for
fitting purposes, we inflated the error bars $\sigma_{v,j}$ of all the
2006~Aug~21 velocities to 3~m~s$^{-1}$. Apparently, for HD~189733,
most of the velocity noise occurs on a time scale longer than one
night.

Our free parameters were the two bodies' masses and radii ($M_S$,
$M_P$, $R_S$ and $R_P$); the orbital inclination ($i$); the
mid-transit time ($T_c$); the line-of-sight stellar rotation velocity
($v \sin i_S$); the angle between the projected stellar spin axis and
orbit normal ($\lambda$; see Ohta et al.~2005 or Gaudi \& Winn~2006
for a diagram of the coordinate system); the velocity zero point
($\gamma$); a velocity offset specific to the night of 2006~Aug~21
($\Delta\gamma$); and a long-term velocity gradient
$\dot{\gamma}$. The parameter $\Delta\gamma$ is needed because of the
photospheric jitter; the first {\it a priori} constraint in Eq.~(2)
enforces a reasonable level of noise. The gradient $\dot{\gamma}$ was
included to account for the long-period orbit of HD~189733 and its
companion star (Bakos et al.~2006a) or possible long-period
planets. We fixed the orbital period to be $2.218575$~days (Bouchy et
al.~2005, H{\'e}brard \& Lecavelier Des Etangs~2006). A well-known
degeneracy prevents $M_S$, $R_S$, and $R_P$ from being determined
independently. We broke this degeneracy with the second {\it a priori}
constraint in Eq.~(2), which enforces the spectroscopic determination
of $M_S$ by Bouchy et al.~(2005).

We used a Markov Chain Monte Carlo algorithm to solve for the model
parameters and their uncertainties (see, e.g., Tegmark et al.~2004 or
Ford 2005). Our jump function was the addition of a Gaussian random
number to each parameter value. We set the perturbation sizes such
that $\sim$20\% of jumps are executed. We created 10 independent
chains, each with 500,000 points, starting from random initial
positions, and discarded the first 20\% of the points in each
chain. The Gelman \& Rubin~(1992) $R$ statistic was close to unity for
each parameter, a sign of good mixing and convergence. We merged the
chains and took the median value of each parameter to be our best
estimate, and the standard deviation as the 1~$\sigma$
uncertainty. For the special case of $v\sin i_S$, we added an
additional error of 6\% in quadrature, due to the systematic error
noted previously.

\section{Results}

The results are given in Table~2. Those parameters depending chiefly
on the photometry ($R_P$, $R_S$, $i$) are in agreement with the most
accurate results reported previously (Bakos et al.~2006b). Likewise,
our result for the planetary mass, $M_P = 1.13\pm 0.03$~$M_{\rm Jup}$,
agrees with the value $1.15\pm 0.04$~$M_{\rm Jup}$ measured by Bouchy
et al.~(2005). Our result for the projected rotation velocity is
$v\sin i_S = 3.0\pm 0.2$~km~s$^{-1}$. This agrees with the value
$3.5\pm 1.0$~km~s$^{-1}$ reported by Bouchy et al.~(2005), which was
based on the line broadening in their disk-integrated stellar
spectrum. It also agrees with the value $3.2\pm 0.7$~km~s$^{-1}$ based
on a similar analysis of our own Keck spectra (D.~Fischer, private
communication). The most interesting result is $\lam$. The sky
projections of the stellar spin axis and the orbit normal are aligned
to within a few degrees.

\section{Summary and Discussion}

We have monitored the apparent Doppler shift of HD~189733 during a
transit of its giant planet. By modeling the RM effect, we find that
the stellar spin axis and the orbit normal are aligned to within a few
degrees.

This is the third exoplanetary system (and the third hot Jupiter) for
which it has been possible to measure $\lambda$. The first system was
HD~209458, for which $\lambda=-4\fdg4\pm 1\fdg4$ (Winn et al.~2005;
see also Wittenmyer et al.~2005, who modeled the RM effect but
required $\lambda=0$). The second system was HD~149026 (Wolf et
al.~2006), for which $\lambda = 11\pm 14\arcdeg$. The small observed
values of $\lambda$ suggest that the most common end-state of the
inward migration of a hot Jupiter involves a close alignment.

With only 3 systems, we cannot yet measure the distribution of
$\lambda$, but we can test the hypothesis of random orientations
(i.e., a uniform distribution in $\lambda$). The weighted mean of the
measured values of $|\lambda|$ is 2.6\arcdeg. If we replace the
measured values by random numbers drawn from a uniform distribution,
the probability that the weighted mean\footnote{Here we have assumed
  the uncertainty in $\lambda$ is independent of $\lambda$, a good
  approximation because all 3 systems have an intermediate impact
  parameter (Gaudi \& Winn 2006).} will be this small
is only 0.04\%. Hence we rule out the hypothesis of random
orientations with 99.96\% confidence.

Winn et al.~(2005) argued that tides from the star would not
ordinarily cause alignment within the star's main-sequence
lifetime. There are therefore two basic possibilities: either the
alignment is primordial and was not disturbed by migration, or there
was a different mechanism to damp any initial or induced
misalignment. Among the various theories of hot Jupiter migration,
some would tend to enhance any initial misalignments, and are thereby
constrained by our results. Such scenarios include planet-planet
scattering followed by circularization (Rasio \& Ford 1996,
Weidenschilling \& Marzari 1996), Kozai migration (Wu \& Murray 2003;
Eggenberger et al.~2004), and tidal capture (Gaudi~2003).

The agreement among the three systems studied to date is clear, but
should not discourage future measurements. Obviously, a sample of
three is only barely sufficient to draw a conclusion. And of course,
the discovery of even a single example of a grossly misaligned system
would be of great interest.

\acknowledgments We thank Debra Fischer for running SME on our
spectra, and Scott Gaudi for very helpful discussions. We recognize
and acknowledge the very significant cultural role and reverence that
the summit of Mauna Kea has always had within the indigenous Hawaiian
community. We are most fortunate to have the opportunity to conduct
observations from this mountain.

\begin{deluxetable}{lcccc}
\tabletypesize{\normalsize}
\tablecaption{Radial Velocities of HD~189733\label{tbl:rv}}
\tablewidth{0pt}

\tablehead{
\colhead{JD} & \colhead{Radial Velocity [m~s$^{-1}$]} & \colhead{Measurement Uncertainty [m~s$^{-1}$]}
}

\startdata
$  2452832.881794$ & $  -13.964  $ & $  1.542 $ \\
$  2452898.800937$ & $  186.717  $ & $  1.480 $ \\
$  2453180.918877$ & $  153.226  $ & $  1.609 $ \\
$  2453240.898507$ & $  121.542  $ & $  1.139 $ \\
$  2453303.750382$ & $ -202.928  $ & $  1.244 $ \\
$  2453303.753576$ & $ -203.187  $ & $  1.139 $ \\
$  2453551.963519$ & $ -126.301  $ & $  1.205 $ \\
$  2453693.688738$ & $   15.243  $ & $  1.228 $ \\
$  2453694.690856$ & $  -84.160  $ & $  1.180 $ \\
$  2453695.688148$ & $  125.535  $ & $  1.259 $ \\
$  2453696.700914$ & $ -206.994  $ & $  1.302 $ \\
$  2453723.712604$ & $   -2.524  $ & $  1.217 $ \\
$  2453926.035150$ & $  159.489  $ & $  1.193 $ \\
$  2453926.887766$ & $ -141.263  $ & $  1.107 $ \\
$  2453927.929745$ & $   53.300  $ & $  1.157 $ \\
$  2453934.845208$ & $  179.802  $ & $  1.224 $ \\
$  2453968.722037$ & $   57.606  $ & $  0.787 $ \\
$  2453968.724225$ & $   57.726  $ & $  0.802 $ \\
$  2453968.726447$ & $   56.423  $ & $  0.832 $ \\
$  2453968.728750$ & $   56.045  $ & $  0.755 $ \\
$  2453968.731227$ & $   54.711  $ & $  0.848 $ \\
$  2453968.733611$ & $   52.911  $ & $  0.798 $ \\
$  2453968.736007$ & $   50.628  $ & $  0.789 $ \\
$  2453968.757928$ & $   42.772  $ & $  0.726 $ \\
$  2453968.760266$ & $   38.559  $ & $  0.760 $ \\
$  2453968.762639$ & $   36.305  $ & $  0.746 $ \\
$  2453968.765023$ & $   37.873  $ & $  0.741 $ \\
$  2453968.767419$ & $   34.160  $ & $  0.818 $ \\
$  2453968.769826$ & $   33.982  $ & $  0.768 $ \\
$  2453968.772280$ & $   32.973  $ & $  0.657 $ \\
$  2453968.774757$ & $   30.196  $ & $  0.740 $ \\
$  2453968.777245$ & $   29.968  $ & $  0.662 $ \\
$  2453968.779688$ & $   28.832  $ & $  0.651 $ \\
$  2453968.782095$ & $   28.517  $ & $  0.764 $ \\
$  2453968.784525$ & $   27.562  $ & $  0.772 $ \\
$  2453968.786979$ & $   26.713  $ & $  0.682 $ \\
$  2453968.789433$ & $   25.029  $ & $  0.682 $ \\
$  2453968.791910$ & $   24.544  $ & $  0.749 $ \\
$  2453968.794433$ & $   20.275  $ & $  0.661 $ \\
$  2453968.796944$ & $   21.667  $ & $  0.855 $ \\
$  2453968.799456$ & $   28.013  $ & $  0.803 $ \\
$  2453968.802049$ & $   31.972  $ & $  0.780 $ \\
$  2453968.804745$ & $   39.982  $ & $  0.797 $ \\
$  2453968.807373$ & $   45.933  $ & $  0.816 $ \\
$  2453968.810000$ & $   47.080  $ & $  0.857 $ \\
$  2453968.812650$ & $   49.130  $ & $  0.870 $ \\
$  2453968.815289$ & $   45.655  $ & $  0.982 $ \\
$  2453968.818021$ & $   42.260  $ & $  0.994 $ \\
$  2453968.820718$ & $   34.659  $ & $  0.965 $ \\
$  2453968.823391$ & $   27.992  $ & $  0.987 $ \\
$  2453968.826134$ & $   18.132  $ & $  0.905 $ \\
$  2453968.828912$ & $    7.218  $ & $  0.858 $ \\
$  2453968.831806$ & $   -0.572  $ & $  1.007 $ \\
$  2453968.834687$ & $   -5.327  $ & $  0.846 $ \\
$  2453968.839641$ & $  -24.613  $ & $  0.751 $ \\
$  2453968.842905$ & $  -34.630  $ & $  0.885 $ \\
$  2453968.845845$ & $  -41.573  $ & $  0.935 $ \\
$  2453968.848877$ & $  -45.183  $ & $  0.842 $ \\
$  2453968.851991$ & $  -47.586  $ & $  0.972 $ \\
$  2453968.855220$ & $  -47.746  $ & $  1.009 $ \\
$  2453968.858356$ & $  -44.215  $ & $  0.928 $ \\
$  2453968.861354$ & $  -37.956  $ & $  0.820 $ \\
$  2453968.864352$ & $  -30.035  $ & $  0.943 $ \\
$  2453968.867361$ & $  -27.041  $ & $  0.890 $ \\
$  2453968.870301$ & $  -25.215  $ & $  0.985 $ \\
$  2453968.873183$ & $  -24.080  $ & $  0.961 $ \\
$  2453968.876076$ & $  -22.205  $ & $  0.892 $ \\
$  2453968.878912$ & $  -27.297  $ & $  0.866 $ \\
$  2453968.881782$ & $  -28.557  $ & $  0.907 $ \\
$  2453968.884583$ & $  -29.194  $ & $  0.822 $ \\
$  2453968.887569$ & $  -33.236  $ & $  0.807 $ \\
$  2453968.890498$ & $  -33.070  $ & $  0.638 $ \\
$  2453968.893403$ & $  -31.439  $ & $  0.847 $ \\
$  2453968.896331$ & $  -36.218  $ & $  0.712 $ \\
$  2453968.899271$ & $  -39.051  $ & $  0.757 $ \\
$  2453968.902118$ & $  -41.663  $ & $  0.900 $ \\
$  2453968.904931$ & $  -42.919  $ & $  0.748 $ \\
$  2453968.907639$ & $  -45.150  $ & $  0.876 $ \\
$  2453968.910278$ & $  -42.636  $ & $  0.956 $ \\
$  2453968.912940$ & $  -43.977  $ & $  0.818 $ \\
$  2453968.915637$ & $  -47.332  $ & $  0.828 $ \\
$  2453968.933565$ & $  -58.550  $ & $  0.715 $ \\
$  2453968.955347$ & $  -69.951  $ & $  0.771 $ \\
$  2453968.977743$ & $  -79.481  $ & $  0.754 $ \\
$  2453969.002674$ & $  -95.622  $ & $  0.750 $ \\
$  2453969.032870$ & $ -112.396  $ & $  0.898 $
\enddata 

\tablecomments{Column 1 gives the Julian Date at the time of the
  photon-weighted mid-exposure. Column 3 gives the measurement
  uncertainties, which do {\it not} include the estimated photospheric
  jitter (see \S~3).}

\end{deluxetable}

\begin{deluxetable}{lcccc}
\tabletypesize{\normalsize}
\tablecaption{System Parameters of HD~189733\label{tbl:params}}
\tablewidth{0pt}

\tablehead{
\colhead{Parameter} & \colhead{Value} & \colhead{Uncertainty}
}

\startdata
                                               $M_S/M_\odot$& $          0.82$  & $           0.03$ \\
                                           $M_P/M_{\rm Jup}$& $          1.13$  & $           0.03$ \\
                                               $R_S/R_\odot$& $          0.73$  & $           0.02$ \\
                                           $R_P/R_{\rm Jup}$& $          1.10$  & $           0.03$ \\
                                                   $i$~[deg]& $          86.1$  & $            0.2$ \\
                                                 $T_c$~[HJD]& $  2453937.7759$  & $         0.0001$ \\
                                  $v \sin i_S$~[km~s$^{-1}$]& $          2.97$  & $           0.22$ \\
                                             $\lambda$~[deg]& $          -1.4$  & $            1.1$ \\
                                       $\gamma$~[m~s$^{-1}$]& $           5.0$  & $           10.1$ \\
                                 $\Delta\gamma$~[m~s$^{-1}$]& $         -15.0$  & $            4.8$ \\
                       $\dot{\gamma}$~[m~s$^{-1}$~yr$^{-1}$]& $          -1.9$  & $            3.3$
\enddata

\end{deluxetable}

\end{document}